\documentclass[11pt,centertags,oneside]{amsart}
\usepackage{amssymb}
\usepackage{ifpdf}

\usepackage[mathcal]{euscript}

\ifpdf
\usepackage[pdftex]{graphicx}
\fi

\usepackage{times}

\newcommand{\Z}{\mathbb{Z}}
\newcommand{\R}{\mathbb{R}}
\newcommand{\C}{\mathbb{C}}
\newcommand{\N}{\mathbb{N}}

\newcommand{\PP}{\mathbb{P}}

\newcommand{\EE}{\mathsf{E}}
\newcommand{\cB}{\mathcal{B}}

\newcommand{\cH}{\mathcal{H}}

\newcommand{\cJ}{\mathcal{J}}

\newcommand{\spec}{{\ensuremath{\rm spec}}}
\newcommand{\specac}{\spec_{\mathrm{ac}}}

\newcommand{\Ran}{\mathrm{Ran}}

\DeclareSymbolFont{ER}{U}{eur}{m}{n}
\DeclareSymbolFont{SY}{U}{psy}{m}{n}

\DeclareMathSymbol{\emptyset}{\mathord}{SY}{'306}
\DeclareMathSymbol{\oplus}{\mathord}{SY}{'305}

\renewcommand{\Im}{{\ensuremath{\mathrm{Im\,}}}}
\renewcommand{\Re}{{\ensuremath{\mathrm{Re\,}}}}
\newcommand{\supp}{{\ensuremath{\mathrm{supp\,}}}}
\newcommand{\tr}{{\ensuremath{\mathrm{tr\,}}}}

\newtheorem{theorem}{Theorem}[section]{\bf}{\it}
\newtheorem{proposition}[theorem]{Proposition}{\bf}{\it}
{\bf}{\it}
{\it}{\rm}
{\bf}{\it}
{\it}{\rm}
{\bf}{\it}

\title[A Random Necklace Model]{A Random Necklace Model}

\author[V.~Kostrykin]{Vadim Kostrykin}
\address{Vadim Kostrykin\\ Fraunhofer-Institut f\"{u}r
Lasertechnik, Steinbachstra{\ss}e 15, D-52074\\ Aachen, Germany}
\email{kostrykin@ilt.fraunhofer.de, kostrykin@t-online.de}

\author[R.~Schrader]{Robert Schrader$^\ast$}
\address{Robert Schrader\\ Institut f\"{u}r Theoretische Physik\\
Freie Universit\"{a}t Berlin, Arnimallee 14\\ D-14195 Berlin, Germany}
\email{schrader@physik.fu-berlin.de}
\thanks{$^\ast$ Supported in part by DFG SFB
288 ``Differentialgeometrie und Quantenphysik''\\ \indent To be published in \textit{Waves in
Random Media}}


\dedicatory{Dedicated to Elliott Lieb on occasion of his 70-th birthday}

\begin{document}

\begin{abstract}
We consider a Laplace operator on a random graph consisting of infinitely
many loops joined symmetrically by intervals of unit length. The arc
lengths of the loops are considered to be independent, identically
distributed random variables. The integrated density of states of this
Laplace operator is shown to have discontinuities provided that the
distribution of arc lengths of the loops has a nontrivial pure point part.
Some numerical illustrations are also presented.
\end{abstract}

\maketitle

\section{Introduction}

In this paper we study a model which perhaps provides the simplest example of a differential
operator on a nontrivial infinite random metric graph. The graph consists of infinitely many loops
joined symmetrically by intervals of unit length. The arc lengths of the loops are independent,
identically distributed random variables. A similar model where the arc lengths of the loops are
kept fixed, was considered by Avron, Exner, and Last in \cite{Avron:Exner:Last}. They called this
model a Necklace of Rings. Mimicking this terminology we will use the name Random Necklace for the
model considered here.

The main objective of the present work is to study the integrated density of states and the
Lyapunov exponent for the Random Necklace Model. It is well known that the integrated density of
states for metrically transitive Schr\"{o}dinger operators on the line is continuous at all energies
since the multiplicity of their spectrum is not greater than two. In contrast, the Laplacian of
the Random Necklace Model can have eigenvalues of infinite multiplicity. Therefore, the integrated
density of states may have discontinuities at those energies. We will show that this is indeed the
case provided that the distribution of arc lengths of the loops has a nontrivial pure point part.
An explicit description of the set of all energies where the integrated density of states is
discontinuous will also be given. Moreover, we will show that the perturbation of the Laplacian by
a magnetic field in general smoothes out the integrated density of states such that some of its
discontinuities disappear.

Discontinuities of the integrated density of states for some discrete
random Laplacians on Delone sets have been studied recently by Klassert,
Lenz, and Stollmann in \cite{Klassert:Lenz:Stollmann}. Earlier such
discontinuities had been observed in \cite{Arai:Tokihiro:Fujiwara:Kohmoto},
\cite{Fujiwara:Arai:Tokihiro:Kohmoto}, \cite{Kohmoto:Sutherland},
\cite{Krajci:Fujiwara} for discrete Laplacians associated with Penrose
tilings. The appearance of discontinuities is again related to the
existence of infinitely degenerated eigenvalues.

The plan of the present work is as follows. The model is defined in Section \ref{Necklace}. In
Section \ref{discont:IDS} we decompose the integrated density of states $N(E)$ into the integrated
density of loop states $N^{\mathrm{loop}}(E)$ (i.e., the states supported on loops of the graph)
and the remainder $\widetilde{N}(E)$. Next we prove that $N^{\mathrm{loop}}(E)$ has
discontinuities if and only if the distribution measure has a nontrivial pure point part. In
Section \ref{Magnetfeld} we show the smoothing effect of a constant magnetic field. In Sections
\ref{IDS} and \ref{Lyapunov} we accommodate the scattering theoretic method proposed in
\cite{Kostrykin:Schrader:1999a} (see also \cite{Kostrykin:Schrader:1998}) to calculate the
integrated density of states and the Lyapunov exponent for the Random Necklace Model. Using this
approach we prove the positivity of the Lyapunov exponent for almost all energies and study the
set of its zeroes. In Section \ref{cont:IDS} we prove the continuity of $\widetilde{N}(E)$ and
discuss its further regularity properties.

We present also some numerical illustrations performed by the method developed in
\cite{Kostrykin:Schrader:1999a}. Part of the material presented here has previously appeared in
\cite{habil}.

\section{Random Necklace}\label{Necklace}

\begin{figure}[htb]
\ifpdf \centerline{\includegraphics[width=120mm]{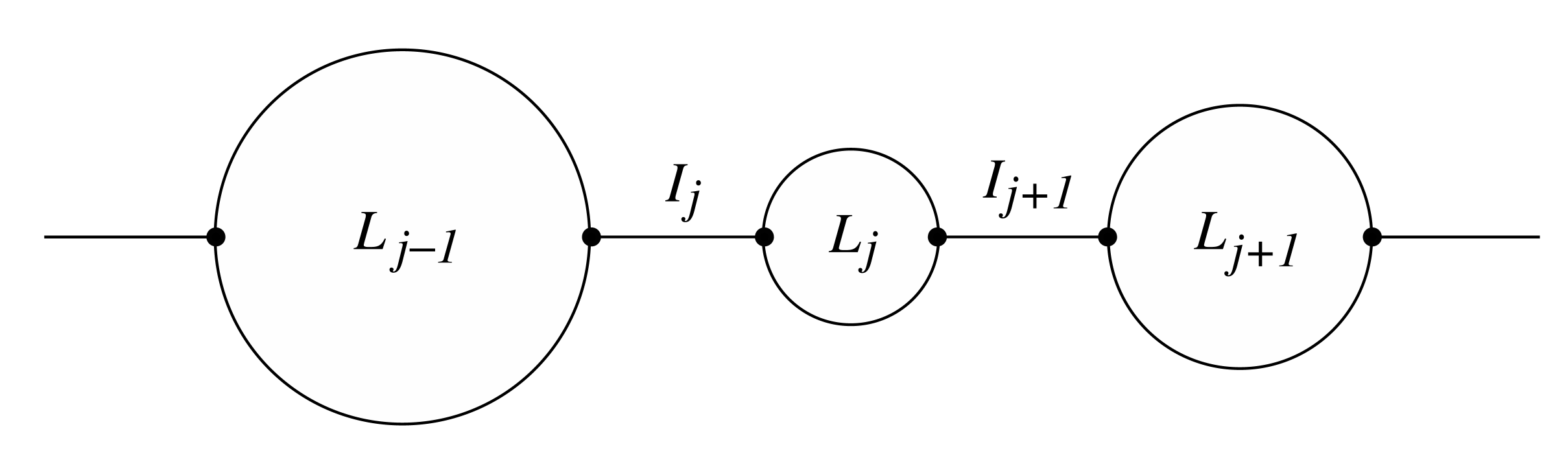}} \else
\vspace{15mm}\centerline{illustration.png}\vspace{15mm} \fi
\caption{\label{fig:ring:illustration} Random necklace.}
\end{figure}

In this section we give a precise formulation of the Random Necklace Model and discuss some of its
basic properties. Consider an infinite graph $\Gamma$ consisting of loops $L_j$ with arc lengths
$2\omega_j$ joined symmetrically by unit intervals $I_j=[0,1]$ (see Fig.\
\ref{fig:ring:illustration}). Any loop will be realized as a union of its upper $L_j^{(+)}\cong
[0,\omega_j]$ and lower $L_j^{(-)}\cong [0,\omega_j]$ parts. We always assume that the left vertex
of any loop $L_j$ corresponds to the point $x=0$ for all three bonds $L_j^{(+)}$, $L_j^{(-)}$, and
$I_j$ adjacent to this vertex. Thus, the right vertex of the loop $L_j$ corresponds to the point
$x=1$ of the interval $I_{j+1}$ and to the point $x=\omega_j$ of the intervals $L_j^{(\pm)}$.
Further, we suppose that $\omega=\{\omega_j\}_{j\in\Z}$ forms an i.i.d.\ sequence of random
variables with distribution measure $\varkappa$ and satisfying $0<\omega_j\leq K$. The underlying
probability space is, therefore, $\Omega=[0,K]^\Z$ with the product probability measure
$\PP=\times_{j\in\Z} \varkappa$.

With the graph $\Gamma$ we associate the Hilbert space $\cH$,
\begin{equation}\label{decomp}
\cH = \bigoplus_{j\in\Z} L^2(I_j)\oplus L^2(L_j)\qquad\textrm{with}\qquad L^2(L_j)=
L^2(L_j^{(+)})\oplus L^2(L_j^{(-)}).
\end{equation}
According to the decomposition \eqref{decomp} we will write the elements $\psi$ of $\cH$ as
follows
\begin{equation*}
\psi = \bigoplus_{j\in\Z} \psi_j^{(0)} \oplus \psi_j^{(+)} \oplus \psi_j^{(-)},
\end{equation*}
where $\psi_j^{(0)}\in L^2(I_j)$ and $\psi_j^{(\pm)}\in L^2(L_j^{(\pm)})$.

Note that the Hilbert space $\cH$ depends on the random variable $\omega=\{\omega_j\}_{j\in\Z}$.
By proper scaling it is possible to formulate the problem equivalently on a Hilbert space
independent of $\omega$ and work with a random operator on a deterministic graph. However, we will
not use this formulation.

On $\cH$ we consider the negative Laplacian $-\Delta(\omega)$ as the operator of second derivative
with local boundary conditions of the form
\begin{equation}\label{rb1}
A\begin{pmatrix} \psi_j^{(0)}(0)\\ \psi_j^{(+)}(0) \\
\psi_j^{(-)}(0) \end{pmatrix} + B\begin{pmatrix} {\psi_j^{(0)}}'(0)\\
{\psi_j^{(+)}}'(0) \\ {\psi_j^{(-)}}'(0)
\end{pmatrix} = 0,\qquad j\in\Z
\end{equation}
at the left vertex of the loop $L_j$ and
\begin{equation}\label{rb2}
A\begin{pmatrix} \psi_j^{(0)}(1)\\ \psi_j^{(+)}(\omega_j) \\
\psi_j^{(-)}(\omega_j)
\end{pmatrix} + B\begin{pmatrix} -{\psi_j^{(0)}}'(1)\\ -{\psi_j^{(+)}}'(\omega_j) \\
-{\psi_j^{(-)}}'(\omega_j)
\end{pmatrix} = 0,\qquad j\in\Z
\end{equation}
at the right vertex. Here $A$ and $B$ are such that they define so called standard boundary
conditions,
\begin{equation}\label{bound.cond}
A=\begin{pmatrix}
        1 & -1 & 0 \\
        0 &  1 & -1 \\
        0 &  0 & 0
        \end{pmatrix}\qquad\textrm{and}\qquad
B=\begin{pmatrix}
        0 & 0 & 0 \\
        0 &  0 & 0\\
        1 &  1 & 1
        \end{pmatrix}.
\end{equation}

By a general result in \cite{Kostrykin:Schrader:1999b} the operator $-\Delta(\omega)$ is
self-adjoint and nonnegative. By the general theory of metrically transitive operators (see, e.g.,
\cite{Pastur:Figotin}) the spectrum of $-\Delta(\omega)$ as well as its components are
deterministic.

It is easy to see that the numbers $E_{jl}=(\pi l /\omega_j)^2$, $j,l\in\Z$
are eigenvalues of $-\Delta(\omega)$ with the eigenfunctions
\begin{equation}\label{eigenf}
\psi_j^{(\pm)}(x)= \pm \sin(\sqrt{E}x)\qquad\textrm{and}\qquad
\psi_k^{(0)}(x)= 0\quad \textrm{for all}\quad k\in\Z.
\end{equation}
Each of these eigenfunctions is compactly supported on the $j$-th loop. Therefore, we will call
these eigenvalues $E_{jl}$ the \emph{loop eigenvalues}. In fact, $-\Delta(\omega)$ has no other
eigenvalues with compactly supported eigenfunctions. The fact that there are no other eigenvalues
with eigenfunctions supported on a single loop can be verified directly. Moreover, in Section
\ref{IDS} below we will prove the following proposition.

\begin{proposition}\label{prop:neq:0}
Let $\psi$ be an arbitrary solution of the Schr\"{o}dinger equation $H \psi = E\psi$. Assume that
$\psi|_{I_j}\neq 0$ on an open subset of some interval $I_j$. Then $\psi\neq 0$ in almost all
internal points of every interval $I_k$, $k\in\Z$.
\end{proposition}

\section{Discontinuities of the Integrated Density of States}\label{discont:IDS}

Let $\Gamma^{m,n}$ with $m,n\in\Z$, $-m\leq n$ be the graph consisting of $n+m+1$ loops $L_j$ of
arc length $\omega_j$, $j=-m,\ldots,n$ joined symmetrically by intervals of unit length. Let
$-\Delta^{m,n}$ be minus the Laplacian for the graph on $\Gamma^{m,n}$ with the boundary condition
\eqref{bound.cond} at all vertices except those vertices of the loops $L_{-m}$ and $L_n$, where no
intervals are attached. At those vertices we impose Dirichlet boundary conditions.

Obviously, the operator $-\Delta^{m,n}$ has a discrete spectrum. Therefore
the finite volume integrated density of states is well-defined:
\begin{equation*}
N^{m,n}(E) := \frac{\tr \EE_{-\Delta^{m,n}}((-\infty,E))}{m+n+1},
\end{equation*}
where $\EE_{-\Delta^{m,n}}(\delta)$ denotes the spectral projection of the operator
$-\Delta^{m,n}$ corresponding to a Borel subset $\delta\subset\R$. By standard arguments (see,
e.g., \cite{Kirsch:Martinelli}) one can prove that the limit
\begin{equation*}
\lim_{m,n\rightarrow\infty}N^{m,n}(E)=:N(E)
\end{equation*}
exists almost surely for all points of continuity of $N(E)$. This limit is deterministic and is
called the integrated density of states. At possible (at most countably many) discontinuity points
of $N(E)$ we make the convention $N(E)=N(E-0)$.

Let $P$ denote the orthogonal projection in $\cH$ onto the subspace generated by the
eigenfunctions corresponding to loop eigenvalues. Then we can decompose the integrated density of
states into two parts
\begin{equation}\label{Schleife}
N^{\mathrm{loop}}(E)=\lim_{m,n\rightarrow\infty}\frac{\tr P
\EE_{-\Delta^{m,n}(\omega)}((-\infty,E))}{m+n+1}
\end{equation}
and
\begin{equation}\label{Schlange}
\widetilde{N}(E)=\lim_{m,n\rightarrow\infty}\frac{\tr (I-P)
\EE_{-\Delta^{m,n}(\omega)}((-\infty,E))}{m+n+1}
\end{equation}
such that $N(E)=\widetilde{N}(E) + N^{\mathrm{loop}}(E)$. It is quite easy to calculate
$N^{\mathrm{loop}}(E)$ explicitly. Indeed, for $E>0$ we have
\begin{equation*}
\begin{split}
N^{\mathrm{loop}}(E) & =\lim_{m,n\rightarrow\infty}
(m+n+1)^{-1}\sum_{j=-m}^n \#\{E_{jl}|\ E_{jl} < E\}\\  &=
\lim_{m,n\rightarrow\infty} (m+n+1)^{-1}\sum_{j=-m}^n
\left\lceil\frac{\omega_j \sqrt{E}}{\pi}\right\rceil,
\end{split}
\end{equation*}
where $\lceil t\rceil$ denotes the largest integer strictly smaller than $t$, $\lceil t\rceil<t$.
By the Birkhoff ergodic theorem the limit exists almost surely and is equal to
\begin{equation}\label{N:loop:integr}
N^{\mathrm{loop}}(E) = \int_0^\infty \left\lceil\frac{\omega_0
\sqrt{E}}{\pi}\right\rceil d\varkappa(\omega_0).
\end{equation}
{}From \eqref{N:loop:integr} it follows that $N^{\mathrm{loop}}(E)=0$ for all $E\leq \pi^2/K^2$,
where $K$ is the suppremum of the topological support of the measure $\varkappa$.

We have the following elementary result on the integrated density of loop
states.

\begin{proposition}\label{prop:3.2}
Let $E \geq \pi^2/K^2$. For all sufficiently small $\varepsilon>0$
\begin{equation*}
N^{\mathrm{loop}}(E+\varepsilon) - N^{\mathrm{loop}}(E)
=\varkappa(S_\varepsilon),
\end{equation*}
where
\begin{equation*}
S_\varepsilon = \bigcup_{k=1}^{\lfloor
K\sqrt{E}/\pi\rfloor}\left(\frac{k\pi}{\sqrt{E+\epsilon}},
\frac{k\pi}{\sqrt{E}}\right]
\end{equation*}
and $\lfloor t \rfloor$ denotes the largest integer not exceeding $t$, $\lfloor t \rfloor \leq t$.
\end{proposition}

The Lebesgue measure of the set $S_\varepsilon$ is obviously bounded by
\begin{equation*}
\varepsilon \sum_{k=1}^{\lfloor K\sqrt{E}/\pi\rfloor}\frac{k\pi}{2E^{3/2}}
\leq \frac{\varepsilon}{4 E} (1+K\sqrt{E}/\pi).
\end{equation*}
Therefore, Proposition \ref{prop:3.2} implies that if the measure
$\varkappa$ is purely continuous, then $N^{\mathrm{loop}}(E)$ is
continuous. Moreover, if $\varkappa$ is purely absolutely continuous with
bounded density, then $N^{\mathrm{loop}}(E)$ is Lipschitz continuous. On
the other hand, one can construct purely absolutely continuous
$\varkappa$'s with \emph{unbounded} density such that the integrated
density of loop states is not H\"{o}lder continuous of any prescribed order.
For instance, fix some $\alpha\in (0,1)$. If $d\varkappa(\omega_0)=\sum_i
c_i |\omega_0-t_i|^{-\alpha} d\omega_0$ with $t_i$ being dense in $(0,K)$
and $c_i>0$ satisfying $\sum_i c_i <\infty$, then $N^{\mathrm{loop}}(E)$ is
H\"{o}lder continuous of order $1-\alpha$ and not of order $\beta>1-\alpha$.
Also, there are singular continuous $\varkappa$'s such that the integrated
density of loop states is not H\"{o}lder continuous of any order $\alpha>0$.


\begin{proof}[Proof of Proposition \ref{prop:3.2}]
For given $E>0$ choose $\varepsilon>0$ so small that
\begin{equation}\label{klein}
\varepsilon < r(E):=\frac{\pi}{K}\left(\frac{\pi}{K} + 2\sqrt{E}\right).
\end{equation}
Then
\begin{equation*}
\frac{K}{\pi}(\sqrt{E+\varepsilon} - \sqrt{E}) < 1
\end{equation*}
and therefore $\left\lceil \frac{\omega_0 \sqrt{E+\varepsilon}}{\pi}\right\rceil - \left\lceil
\frac{\omega_0 \sqrt{E}}{\pi}\right\rceil \leq 1$ for any $0<\omega_0\leq K$. The equality sign
occurs if and only if
\begin{equation*}
\frac{\omega_0 \sqrt{E+\varepsilon}}{\pi} > k \geq  \frac{\omega_0 \sqrt{E}}{\pi}
\end{equation*}
for some integer $k$. Thus,
\begin{equation*}
\left\lceil \frac{\omega_0 \sqrt{E+\varepsilon}}{\pi}\right\rceil - \left\lceil \frac{\omega_0
\sqrt{E}}{\pi}\right\rceil = \chi_{S_\varepsilon}(\omega_0),
\end{equation*}
where $\chi_{S_\varepsilon}$ is the characteristic function of the set
$S_\varepsilon$. The claim follows from equation \eqref{N:loop:integr}.
\end{proof}


\begin{theorem}\label{thm:1}
Assume that the probability measure $\varkappa$ has a nontrivial pure point
part $\varkappa_{\mathrm{pp}}$,
\begin{equation}\label{pure:point}
\varkappa_{\mathrm{pp}}(\cdot) = \sum_{i=1}^\infty p_i \delta_{s_i}(\cdot)
\qquad\textrm{with}\qquad p_i\geq 0,\qquad 0 < \sum_{i=1}^\infty p_i \leq 1,
\end{equation}
where $\delta_{s_i}$ denotes the Dirac measure concentrated at $s_i$.
Consider the nonempty set
\begin{equation}\label{D:kappa}
D_\varkappa:=\{E\in\R_+|\ E=(\pi k/s_i)^2, \quad s_i\neq 0, \quad p_i\neq
0, \quad k\in\N\}.
\end{equation}
Then $D_\varkappa$ is the set of all points of discontinuity of the
integrated density of states $N(E)$ and for any $E\in D_\varkappa$
\begin{equation}\label{equality}
N(E+0)-N(E) = \sum_{i=1}^\infty \alpha_i p_i > 0
\end{equation}
with $\alpha_i=1$ if $s_i \sqrt{E}/\pi$ is integer and $\alpha_i=0$
otherwise.
\end{theorem}


\begin{proof} {}From Proposition \ref{prop:3.2} it follows that
\begin{equation*}
N^{\mathrm{loop}}(E+0) - N^{\mathrm{loop}}(E) = \varkappa(M),
\end{equation*}
where $M=\{\omega_0|\ \omega_0 \sqrt{E}/\pi\ \textrm{is integer}\}$. Obviously, for any $E$ the
set $M$ is at most discrete. Therefore, we have
\begin{equation}\label{equality:2}
N^{\mathrm{loop}}(E+0)-N^{\mathrm{loop}}(E) = \varkappa_{\mathrm{pp}}(M) =
\sum_{i=1}^\infty \alpha_i p_i \neq 0
\end{equation}
if $E\in D_\varkappa$ and zero otherwise.

Since $\widetilde{N}(E)$ is non-decreasing it follows that $N(E)$ is discontinuous on
$D_\varkappa$. Moreover, we obtain \eqref{equality} with ``$=$" being replaced by ``$\geq$". Thus,
to complete the proof it suffices to show that $\widetilde{N}(E)$ defined by \eqref{Schlange} is
continuous. We will prove this fact in Section \ref{cont:IDS}.
\end{proof}


Since the set $M$ is discrete the sum in equation \eqref{equality} involves only a finite number
of nonvanishing terms.

If the distribution measure $\varkappa$ has a nontrivial singular
continuous component, then the canonical decomposition of
$N^{\mathrm{loop}}(E)$ contains a nontrivial singular continuous part. This
is established in the following proposition.

\begin{proposition}\label{prop:neu}
If $\varkappa$ is purely singular continuous, then $N^{\mathrm{loop}}(E)$
is a singular continuous function.
\end{proposition}


\begin{proof} Since $N^{\mathrm{loop}}(E)$ is non-decreasing, it is
differentiable for Lebesgue almost all $E\in\R$. By Theorem \ref{thm:1} it
is continuous. Therefore, to prove the claim it suffices to show that $d
N^{\mathrm{loop}}(E)/dE=0$ for a.e.\ $E$.

Fix an arbitrary $E_0\geq \pi^2/K^2$ and consider
\begin{equation*}
F_k(\varepsilon) = \varkappa\left(\left(\frac{k\pi}{\sqrt{E_0+\varepsilon}},
\frac{k\pi}{\sqrt{E_0}} \right] \right), \qquad k\in\N
\end{equation*}
as a function of $\varepsilon\in (0,r(E_0))$ with $r(E_0)$ being defined in \eqref{klein}. Since
$\varkappa$ is singular continuous, we have $F'_k(\varepsilon)=0$ for a.e.\ $\varepsilon$. {}From
Proposition \ref{prop:3.2} it follows that
\begin{equation*}
N^{\mathrm{loop}}(E_0+\varepsilon) = N^{\mathrm{loop}}(E_0) + \sum_{k=1}^{\lfloor
K\sqrt{E_0}/\pi\rfloor} F_k(\varepsilon)
\end{equation*}
for all $\varepsilon\in (0,r(E_0))$. Thus, for almost all $E\in (E_0, E_0+r(E_0))$ we have
\begin{equation*}
\frac{d N^{\mathrm{loop}}(E)}{dE}=\frac{d
N^{\mathrm{loop}}(E_0+\varepsilon)}{d\varepsilon} = \sum_{k=1}^{\lfloor
K\sqrt{E_0}/\pi\rfloor} F'_k(\varepsilon) = 0.
\end{equation*}
This proves the claim since $E_0$ is arbitrary.
\end{proof}

\section{Magnetic Field Smoothes the Integrated Density of States}\label{Magnetfeld}

Without loss of generality we can assume that the graph $\Gamma$ is imbedded in a plane in $\R^3$
and the loops $L_j$ are circles. In this section we consider a magnetic perturbation
$-\Delta(\omega;\cB)$ of the Laplacian described in Section \ref{Necklace}. Assume there is a
constant magnetic field $\cB$ perpendicular to the plane containing the graph. Since the $j$-th
loop encloses an area $\omega_j^2/\pi$ the magnetic flux through this loop is given by
$\Phi_j=\omega_j^2 \cB/\pi$. As shown in \cite{Kostrykin:Schrader:2003} prescribing magnetic
fluxes through all loops of the graph defines a magnetic Laplacian uniquely up to a gauge
transformation. The resulting magnetic Laplacian is again defined to be the operator of second
derivative but now with different boundary conditions. The boundary condition \eqref{rb1} at the
left vertex of the loop $L_j$ remains unchanged and the boundary condition \eqref{rb2} at the
right vertex takes the form
\begin{equation*}
A_j\begin{pmatrix} \psi_j^{(0)}(1)\\ \psi_j^{(+)}(\omega_j) \\
\psi_j^{(-)}(\omega_j)
\end{pmatrix} + B_j\begin{pmatrix} -{\psi_j^{(0)}}'(1)\\ -{\psi_j^{(+)}}'(\omega_j) \\
-{\psi_j^{(-)}}'(\omega_j)
\end{pmatrix} = 0,\qquad j\in\Z
\end{equation*}
with
\begin{equation}\label{bound:cond:magnet}
A_j=\begin{pmatrix}
        1 & -1 & 0 \\
        0 &  e^{i\Phi_j/2} & -e^{-i\Phi_j/2} \\
        0 &  0 & 0
        \end{pmatrix},\qquad
B_j=\begin{pmatrix}
        0 & 0 & 0 \\
        0 &  0 & 0\\
        1 &  e^{i\Phi_j/2} & e^{-i\Phi_j/2}
        \end{pmatrix}.
\end{equation}

It is easy to see that the numbers $E_{jl}=(\pi l/\omega_j)^2$ are
eigenvalues of $-\Delta(\omega;\cB)$ if and only if $\cB \omega_j^2/\pi^2$
is integer. The corresponding eigenfunctions are given by \eqref{eigenf}.
Also, there are no other eigenvalues with compactly supported
eigenfunctions.

Assume again that the probability measure for the i.i.d.\ distributed arc lengths $\omega_j$ has a
nontrivial pure point part given by \eqref{pure:point} and consider the set
\begin{equation*}
D_\varkappa(\cB):=\{E\in\R_+|\ E=(\pi k/s_i)^2, \quad s_i\neq 0,\quad \cB
s_j^2/\pi^2 \in\N_0, \quad p_i\neq 0, \quad k\in\N\}.
\end{equation*}
Similarly to the analysis of Section \ref{discont:IDS} one can show that $N^{\mathrm{loop}}(E)$ is
discontinuous on the set $D_\varkappa(\cB)\subseteq D_\varkappa$ and nowhere else. Note that for
$\cB\neq 0$ this set is in general strictly smaller than $D_\varkappa$ defined by \eqref{D:kappa}.
Since all discontinuities of the integrated density of states are contained in
$N^{\mathrm{loop}}(E)$ this implies that generically (if $\cB s_j^2/\pi$ is not integer) the
discontinuities disappear under the perturbation by a magnetic field.

\section{The Integrated Density of States and Scattering Amplitudes}\label{IDS}

To proceed further with the analysis of the Laplacian $-\Delta(\omega)$ we will use some results
from scattering theory. Let $\Gamma_{m,n}$ with $m,n\in\Z$, $-m\leq n$ be the graph consisting of
$n+m+1$ loops $L_j$ of arc length $\omega_j$, $j=-m,\ldots,n$ joined symmetrically by the
intervals of unit length and of two semi-lines attached to the loops $L_{-m}$ and $L_n$. With this
graph we associate the Hilbert space $\cH_{m,n}=\cH^{\mathrm{ext}}\oplus\cH^{\mathrm{int}}_{m,n}$,
where $\cH^{\mathrm{ext}}=L^2(0,\infty)\oplus L^2(0,\infty)$ and
\begin{equation*}
\cH^{\mathrm{int}}_{m,n} = \left\{\begin{array}{ll} \displaystyle
L^2(L_{-m})\ \oplus \bigoplus_{j=-m+1}^n
\left[L^2(I_j)\oplus L^2(L_j)\right], & m,n\neq 0, \\
L^2(L_0), & m=n=0. \end{array}\right.
\end{equation*}

By $-\Delta_{m,n}(\omega)$ we denote minus the Laplacian acting on $\cH_{m,n}$ with the boundary
conditions \eqref{bound.cond}. We consider the scattering matrix and the spectral shift function
for the pair of operators $(-\Delta_{m,n}(\omega),-\Delta)$ where $\Delta$ is a usual Laplacian on
$L^2(\R)$. Although these operators act in different Hilbert spaces, the scattering matrix as well
as the spectral shift function can be constructed in this case (see, e.g., \cite{Yafaev} and the
\ref{Appendix:A}).

Identifying in a natural way $L^2(\R)$ and $\cH^{\mathrm{ext}}$ we define the isometric
identification operator $\cJ:\ L^2(\R)\rightarrow\cH$ such that $\Ran \cJ=\cH^{\mathrm{ext}}$.
Obviously, $I-\cJ^\ast\cJ=0$ and $I-\cJ\cJ^\ast=P_{\cH^{\mathrm{ext}}}$ with
$P_{\cH^{\mathrm{ext}}}$ being the projection in $\cH_{m,n}$ onto the subspace
$\cH^{\mathrm{ext}}$. It is easy to check that the conditions \eqref{Koplienko} in the
\ref{Appendix:A} are fulfilled for, e.g., $k=1$. Thus the spectral shift function
$\xi_{m,n}(E;\omega):=\xi(E;-\Delta_{m,n}(\omega),-\Delta;\cJ)$ exists and satisfies the trace
formula \eqref{Koplienko.trace}. The condition $\xi_{m,n}(E;\omega)=0$ for $E<0$ fixes the
spectral shift function uniquely. The scattering matrix
\begin{equation*}
S_{m,n}(E;\omega):=S(-\Delta_{m,n}(\omega),-\Delta;\cJ)=\begin{pmatrix}
T_{m,n}(E;\omega) & R_{m,n}(E;\omega) \\ L_{m,n}(E;\omega) &
T_{m,n}(E;\omega)
\end{pmatrix}
\end{equation*}
is defined as in Section 2 of \cite{Kostrykin:Schrader:2001b}.

The operator $-\Delta_{0,0}(\omega)$ on the graph $\Gamma_{0,0}$ consisting of a single loop and
two half-lines was considered in Example 3.2 in \cite{Kostrykin:Schrader:1999b}, where it was
shown that the transmission and reflection amplitudes are given by
\begin{equation}\label{T.R.L.}
T_{0,0}(E;\omega)=-\frac{8e^{i\omega_0\sqrt{E}}}{e^{2i\omega_0\sqrt{E}}-9},\quad
R_{0,0}(E;\omega)=L_{0,0}(E;\omega)=-\frac{3(e^{2i\omega_0\sqrt{E}}-1)}{e^{2i\omega_0\sqrt{E}}-9}.
\end{equation}
The operator $-\Delta_{0,0}(\omega)$ has infinitely many eigenvalues
$\{\pi^2 n^2/ \omega_0^2,\ n\in\N\}$ imbedded in the absolutely continuous
spectrum. At those energies the reflection coefficients vanish and
$T_{0,0}(\pi^2n^2/a^2;\omega)=(-1)^{n+1}$.

The spectral shift function $\xi_{0,0}(E;\omega)$ can be calculated
explicitly. Indeed, by the chain rule for the spectral shift function we
have
\begin{equation*}
\xi_{0,0}(E;\omega) = \xi(E;
-\Delta_{0,0}(\omega),(-\Delta)\oplus(-\Delta^{\mathrm{loop}})) +\xi(E;
(-\Delta)\oplus(-\Delta^{\mathrm{loop}}),-\Delta;\cJ),
\end{equation*}
where $\Delta^{\mathrm{loop}}$ denotes the (self-adjoint) Laplace operator on
$\cH^{\mathrm{int}}_{0,0}=L^2([0,\omega_0])\oplus L^2([0,\omega_0])$ with the boundary conditions
\begin{equation*}
\begin{split}
\psi^{(+)}(0)=\psi^{(-)}(0),\qquad &
{\psi^{(+)}}^\prime(0)+{\psi^{(-)}}^\prime(0)= 0,\\
\psi^{(+)}(\omega_0)=\psi^{(-)}(\omega_0), \qquad &
{\psi^{(+)}}^\prime(\omega_0)+{\psi^{(-)}}^\prime(\omega_0)=0.
\end{split}
\end{equation*}
Combining the trace formula \eqref{Koplienko.trace} with Birman-Krein
Theorem \eqref{Appendix:Birman:Krein} (both with with $\cJ=I$) we obtain
\begin{equation*}
\xi(E; -\Delta_{0,0}(\omega),(-\Delta)\oplus(-\Delta^{\mathrm{loop}})) =
-\frac{1}{\pi} \phi_{0,0}(E;\omega)
\end{equation*}
with
\begin{equation*}
\begin{split}
\phi_{0,0}(E;\omega) &:= \frac{1}{2i} \log\det S(E;
-\Delta_{0,0}(\omega),(-\Delta)\oplus(-\Delta^{\mathrm{loop}});I)\\ & =
\mathrm{Arctan}\left(\frac{5}{4}\tan(\sqrt{E}\omega_0) \right),
\end{split}
\end{equation*}
where $\mathrm{Arctan}$ is chosen such that $x\mapsto \mathrm{Arctan}(\tan x)$ is continuous for
all $x\in\R$ and $\mathrm{Arctan}(0)=0$. In particular, $\phi_{0,0}(\pi^2k^2/a^2;\omega)$ $=\pi k$
and $\phi_{0,0}(\pi^2(k+1/2)^2/a^2;\omega)=\pi(k+1/2)$ for all $k\in\N_0$ and all
$\omega\in\Omega$. Furthermore,
\begin{equation*}
\xi(E; (-\Delta)\oplus(-\Delta^{\mathrm{loop}}),-\Delta;\cJ)=\xi(E;
(-\Delta)\oplus(-\Delta^{\mathrm{loop}}),-\Delta\oplus 0;I)
\end{equation*}
and, thus, by Lemma 3.1 in \cite{Kostrykin:Schrader:1999a} equals minus the eigenvalue counting
function for the operator $-\Delta^{\mathrm{loop}}$,
\begin{equation*}
\xi(E; (-\Delta)\oplus(-\Delta^{\mathrm{loop}}),-\Delta;\cJ)=-\lceil\sqrt{E}
\omega_0/\pi\rceil.
\end{equation*}
Therefore, we obtain
\begin{equation*}
\xi_{0,0}(E;\omega) = -\frac{1}{\pi}
\mathrm{Arctan}\left(\frac{5}{4}\tan(\sqrt{E}\omega_0) \right) -
\lceil\sqrt{E} \omega_0/\pi\rceil.
\end{equation*}

In a similar way one can calculate the spectral shift function $\xi_{m,n}(E;\omega)$ for arbitrary
integers $m$ and $n$ such that $-m\leq n$,
\begin{equation*}
\xi_{m,n}(E;\omega) = -\frac{1}{\pi} \phi_{m,n}(E;\omega) -
\sum_{j=-m}^n\lceil\sqrt{E} \omega_j/\pi\rceil,
\end{equation*}
where $\phi_{m,n}(E;\omega)$ is the scattering phase for the pair of
operators $(-\Delta_{m,n}(\omega),-\Delta)$.

Denoting by $N_0(E)=\sqrt{E}/\pi$ the integrated density of states for the
Laplacian $-\Delta$ on $L^2(\R)$, by results of
\cite{Kostrykin:Schrader:1999a} (Theorem 4.1, equation (4.4), and Theorem
4.4) we obtain that
\begin{equation}\label{NvonE}
N(E) = N_0(E) -  \lim_{m,n\rightarrow\infty}
\frac{\xi_{m,n}(E;\omega)}{n+m+1}
\end{equation}
almost surely. Although the results of \cite{Kostrykin:Schrader:1999a} are
formulated and proven for Schr\"{o}dinger operators on the line, all proofs
extend verbatim to the present context.

\begin{figure}[htb]
\ifpdf \centerline{\includegraphics[width=120mm]{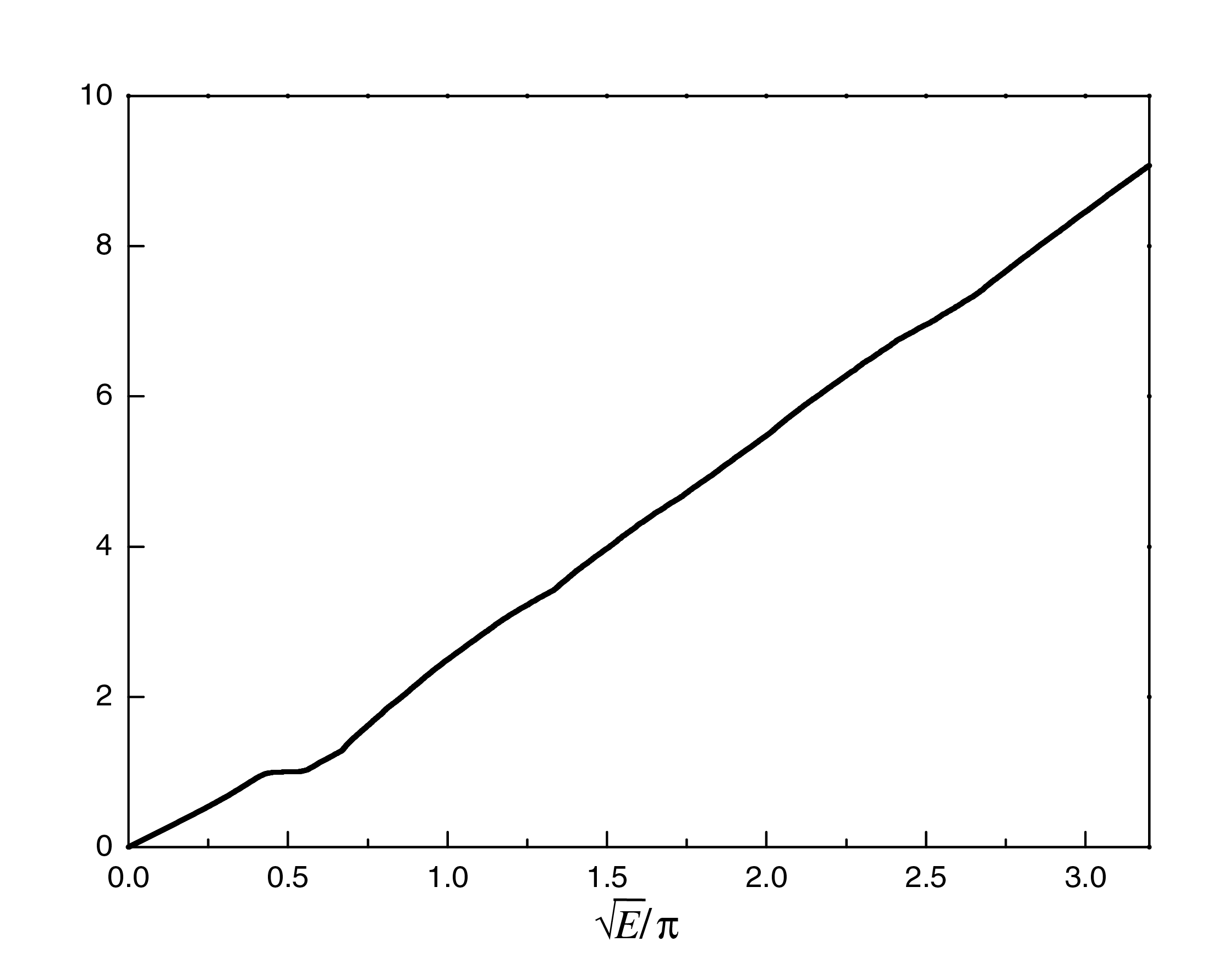}} \else
\vspace{55mm} \centerline{stetig\_N.png} \vspace{55mm} \fi
\caption{\label{fig:con:necklace} The integrated density of states $N(E)$
for the case of uniformly continuously distributed $\omega_0\in[1/2,3/2]$.
The horizontal plateau corresponds to a spectral gap.}
\end{figure}

Let
\begin{equation}\label{das:braucht}
\Lambda_{\omega_0}(E)=\begin{pmatrix}
\frac{e^{-i\sqrt{E}}}{T_{0,0}(E;\omega)} &
-\frac{R_{0,0}(E;\omega)}{T_{0,0}(E;\omega)} \\
\frac{L_{0,0}(E;\omega)}{T_{0,0}(E;\omega)} &
\frac{e^{i\sqrt{E}}}{T_{0,0}(E;\omega)^\ast}
\end{pmatrix}
\end{equation}
and
\begin{equation*}
e_+ = \begin{pmatrix} 1 \\ 0
\end{pmatrix}\qquad\textrm{and}\qquad e_- = \begin{pmatrix} 0 \\
1 \end{pmatrix}.
\end{equation*}
By the arguments presented in \cite{Kostrykin:Schrader:1999a} from equation
\eqref{NvonE} we obtain the following theorem.

\begin{theorem}\label{thm:2}
For $E>0$ the integrated density of states $N(E)$ is given by
\begin{equation}\label{multip}
N(E) = \mp \frac{1}{\pi} \lim_{m,n\rightarrow\infty}
\frac{\varphi_{m,n}^{\pm}(E;\omega)}{m+n+1}  + N^{\mathrm{loop}}(E),
\end{equation}
where
\begin{equation}\label{produkt}
\varphi_{mn}^\pm(E;\omega) := \arg\left\langle e_\pm, \prod_{j=-m}^n
\Lambda_{\omega_j}(E) e_\pm\right\rangle,
\end{equation}
$\langle\cdot,\cdot\rangle$ denotes the inner product in $\C^2$ and $N^{\mathrm{loop}}$ is given
by \eqref{N:loop:integr}. The choice of the argument is uniquely fixed by the condition
\begin{equation*}
\varphi_{m,k}^\pm+\varphi_{-k,n}^\pm-\frac{\pi}{2} \leq \varphi_{mn}^\pm
\leq \varphi_{m,k}^\pm+\varphi_{-k,n}^\pm +\frac{\pi}{2}
\end{equation*}
for arbitrary $k\in\Z$ satisfying $-m\leq k$ and $-k\leq n$.
\end{theorem}

\begin{figure}[htb]
\ifpdf \centerline{\includegraphics[width=120mm]{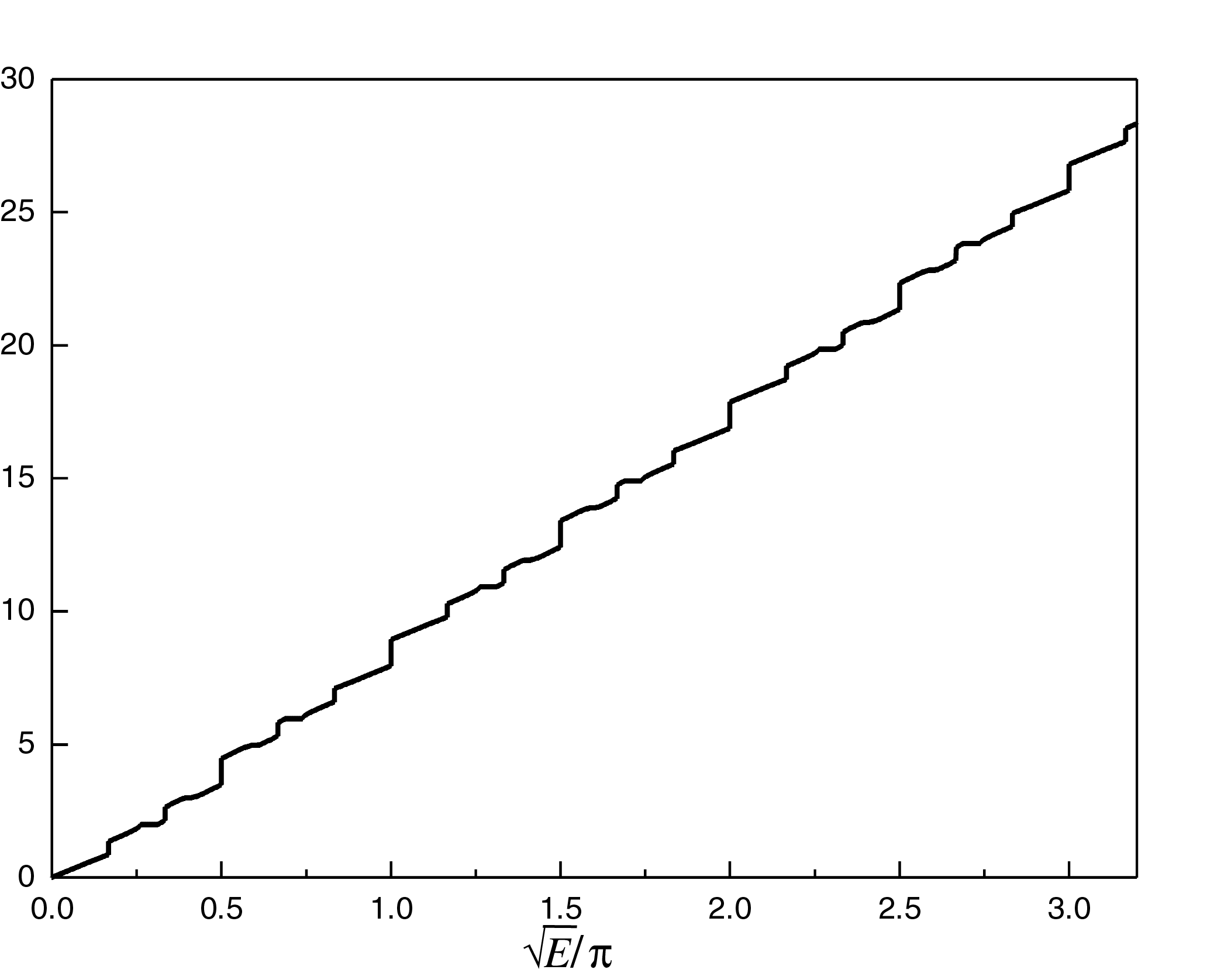}} \else
\vspace{60mm}\centerline{Bernoulli\_N.png} \vspace{60mm}\fi
\caption{\label{fig:Bernoulli:N} The integrated density of states $N(E)$
for the Bernoulli distribution $\varkappa= \frac{1}{2}\delta_2 +
\frac{1}{2}\delta_6$, i.e., $\omega_j\in\{2,6\}$ with equal probability.
Horizontal plateaus correspond to spectral gaps, vertical strokes represent
discontinuities.}
\end{figure}

In equation \eqref{produkt} and below the product $\prod$ is to be understood in the ordered
sense. Theorem \ref{thm:2} provides an algorithm for numerical calculations of the integrated
density of states. Figures \ref{fig:con:necklace} and \ref{fig:Bernoulli:N} show examples of such
calculations. Some other numerical results for Schr\"{o}dinger operators on the line are presented in
\cite{Kostrykin:Schrader:1998}.

By means of Theorem \ref{thm:2} one can obtain (see
\cite{Kostrykin:Schrader:2000b}) a two-sided estimate on the integrated
density of states $\widetilde{N}(E)$
\begin{equation}\label{estimate}
\left|\widetilde{N}(E)-N_0(E) - \frac{1}{\pi} \int_0^\infty
\mathrm{Arctan}\left(\frac{5}{4}\tan(\omega_0\sqrt{E})\right)d\varkappa(\omega_0)\right|
\leq \frac{1}{2}.
\end{equation}

Comparing \eqref{multip} with equations \eqref{Schleife} and
\eqref{Schlange} we obtain that
\begin{equation*}
\widetilde{N}(E) = \mp \frac{1}{\pi} \lim_{m,n\rightarrow\infty}
\frac{\varphi_{m,n}^\pm(E)}{m+n+1}
\end{equation*}
for almost all $\omega\in\Omega$. A simple calculation now leads to the following representation
for $\widetilde{N}(E)$:
\begin{equation}\label{Differenz}
\widetilde{N}(E) = N_0(E) -\frac{1}{\pi} \lim_{m,n\rightarrow\infty}
\frac{\phi_{m,n}(E;\omega)}{m+n+1},
\end{equation}
where $\phi_{m,n}(E;\omega)$ is the scattering phase for the pair of operators
$(-\Delta_{m,n}(E;\omega),-\Delta)$. In the next section we use equation \eqref{Differenz} to
prove the Thouless formula in the present context.

Now we are in position to prove Proposition \ref{prop:neq:0}. Let
$\Lambda_{m,n}(E;\omega)$ denote the transfer matrix,
\begin{equation}\label{Lambda:def}
\Lambda_{m,n}(E;\omega) = \begin{pmatrix} \frac{1}{T_{m,n}(E;\omega)} &
-\frac{R_{m,n}(E;\omega)}{T_{m,n}(E;\omega)}
\\ \frac{L_{m,n}(E;\omega)}{T_{m,n}(E;\omega)} &
\frac{1}{T_{m,n}(E;\omega)^\ast}
\end{pmatrix}.
\end{equation}

\begin{proof}[Proof of Proposition \ref{prop:neq:0}]
Note that $\psi|_{I_k}$ for any $k\in\Z$ is necessarily of the form $a_k
e^{i\sqrt{E}x} + b_k e^{-i\sqrt{E}x}$. Therefore, if $\psi|_{I_j}\neq 0$ on
an open subset of some interval $I_j$, then $|a_j| + |b_j| \neq 0$. For
$k\neq j$ the coefficients $a_k$, $b_k$ are determined by the equation
(see, e.g., \cite{Kostrykin:Schrader:2001b})
\begin{equation}\label{Kette}
\begin{pmatrix} a_k\\ b_k\end{pmatrix} = \Lambda_{-k,j}(E;\omega) \begin{pmatrix} a_j\\
b_j\end{pmatrix},
\end{equation}
where $\det\Lambda_{-k,j}(E,\omega)=1$. Assume that $a_k = b_k=0$ for some
$k\in\Z$. Then from \eqref{Kette} it follows that $a_j=b_j=0$, which is a
contradiction.
\end{proof}


\section{The Lyapunov Exponent}\label{Lyapunov}

Following convention we define the Lyapunov exponent for the Random Necklace
Model as the exponential growth rate of the norm of the transfer matrix,
\begin{equation}\label{def:Lyapunov}
\gamma(E) = \lim_{m,n\rightarrow\infty} \frac{\log\|\Lambda_{m,n}(E,\omega)\|}{n+m+1}.
\end{equation}
It is a general fact that for every $E>0$ this limit exists almost surely and is nonnegative.
Moreover, (see Theorems 5.1 and 5.3 in \cite{Kostrykin:Schrader:1999a}, also cf.\
\cite{Lifshitz:Gredeskul:Pastur:82}, \cite{Lifshitz:Gredeskul:Pastur:88}, \cite{Marchenko:Pastur})
\begin{equation}\label{Lyapunov:2}
\gamma(E)=-\lim_{m,n\rightarrow\infty} \frac{\log\|T_{m,n}(E;\omega)\|}{n+m+1},
\end{equation}
where $T_{m,n}(E;\omega)$ is the transmission coefficient corresponding to the Laplacian
$-\Delta_{m,n}(\omega)$. Also, \eqref{def:Lyapunov} can be rewritten as
\begin{align}\label{multip:gam}
\gamma(E) & =  \lim_{m,n\rightarrow\infty} \frac{1}{m+n+1}
\log\left|\left\langle e_\pm, \prod_{j=-m}^n \Lambda_{\omega_j}(E)
e_\pm\right\rangle\right| \\
&= \lim_{m,n\rightarrow\infty} \frac{1}{m+n+1} \log\left\| \prod_{j=-m}^n
\Lambda_{\omega_j}(E) \right\|,\nonumber
\end{align}
where the matrices $\Lambda_{\omega_j}(E)$ are defined by
\eqref{das:braucht}.

\begin{figure}[htb]
\ifpdf \centerline{\includegraphics[width=120mm]{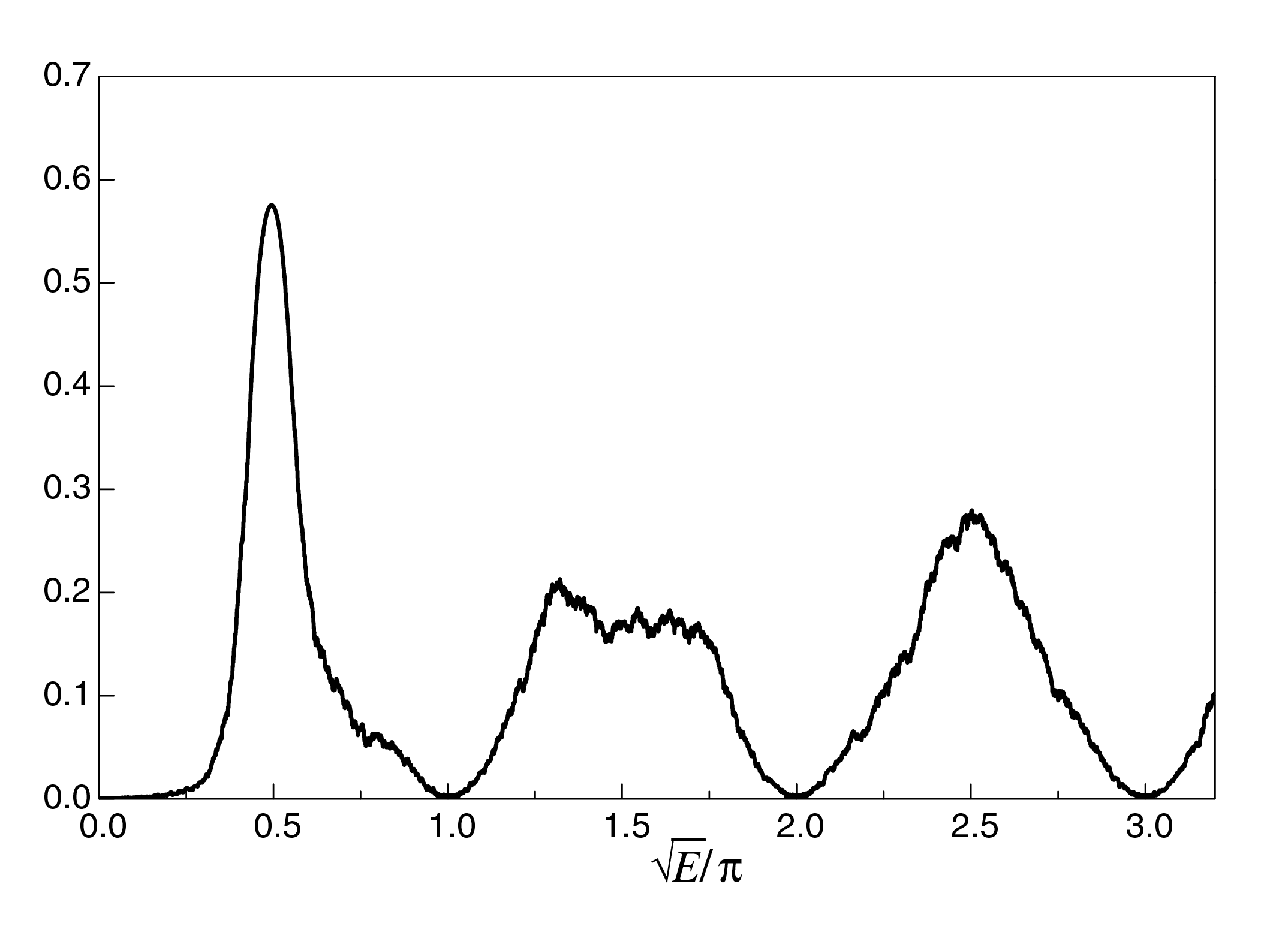}} \else
\vspace{53mm}\centerline{stetig\_gam.png} \vspace{53mm} \fi
\caption{\label{fig:ring:con} The Lyapunov exponent for the case of
uniformly continuously distributed $\omega_0\in[1/2,3/2]$.}
\end{figure}

Kotani's Theorem, which states that the Lyapunov exponent of every Schr\"{o}dinger operator on the
line with non-deterministic potential is almost everywhere positive (see, e.g.,
\cite{Carmona:Lacroix}), does not apply directly to the model considered here. Therefore, to prove
that $\gamma(E)$ is positive for almost all $E>0$ in the case when $\supp\varkappa$ contains at
least one non-isolated point we refer to Theorem 5.6 of \cite{Kostrykin:Schrader:1999a}. By this
result for $E>0$ the Lyapunov exponent vanishes on the set $\{E=(\pi n)^2|\ n\in\N\}$ and nowhere
else. Also, for $E=0$ the matrix ${\Lambda}_{\omega_0}(E)$ equals the unit matrix and hence
$\gamma(0)=0$. As an illustration we have computed the the Lyapunov exponent for uniformly
continuously distributed $\omega_j$'s on the interval $[1/2,3/2]$, i.e., for
$d\varkappa(E)=\chi_{[1/2,3/2]}(E)$ (see Figure \ref{fig:ring:con}).

The case when the measure $\varkappa$ is purely discrete (and thus is a finite convex combination
of Dirac measures) is not covered by Theorem 5.6 in \cite{Kostrykin:Schrader:1999a}. First,
consider the Bernoulli distribution $\varkappa=p\delta_{s_1}+(1-p)\delta_{s_2}$, $0<p<1$ since in
this case we may invoke the recent results of Damanik, Sims, and Stolz \cite{Damanik:Sims:Stolz:1}
(see also \cite{Damanik:Sims:Stolz}). To apply this result we need to introduce the scattering
amplitudes $T^{(s_1,s_0)}$, $R^{(s_1,s_0)}$, $L^{(s_1,s_0)}$ for $-\Delta_{0,0}(s_1)$ relative to
the ``background" operator $-\Delta_{0,0}(s_0)$. (As discussed above the fact that these operators
act in different Hilbert spaces is not relevant). A simple calculation shows that they can be
determined from the relation
\begin{equation*}
\begin{pmatrix} \frac{1}{T^{(s_1,s_0)}(E)} & -\frac{R^{(s_1,s_0)}(E)}{T^{(s_1,s_0)}(E)} \\
\frac{L^{(s_1,s_0)}(E)}{T^{(s_1,s_0)}(E)} &
\frac{1}{T^{(s_1,s_0)}(E)^\ast}\end{pmatrix} = \Lambda_{0,0}(E;s_1)\
\Lambda_{0,0}(E;s_0)^{-1},
\end{equation*}
where $\Lambda_{0,0}$ is defined in \eqref{Lambda:def}. In particular, we
obtain
\begin{equation*}
R^{(s_1,s_0)}(E) = L^{(s_1,s_0)}(E) = -\frac{3(e^{2i s_1 \sqrt{E}}-e^{2i s_0
\sqrt{E}})}{e^{2i s_1 \sqrt{E}}-9e^{2i s_0 \sqrt{E}}}
\end{equation*}
such that the reflection coefficients vanish if and only if $\sqrt{E}(s_1-s_0)/\pi$ is an integer.
Therefore, by applying Theorem 1 of \cite{Damanik:Sims:Stolz:1} we conclude that the Lyapunov
exponent vanishes on the set
\begin{equation}\label{das:S}
S(s_0,s_1):=\left\{E=\left(\frac{\pi k}{2}\right)^2,\ k\in\N\right\}\cup
\left\{E=\left(\frac{\pi k}{s_1-s_0}\right)^2,\ k\in \N\right\}
\end{equation}
and nowhere else.

Let now $\varkappa$ be an arbitrary discrete measure given by \eqref{pure:point} with a finite
number of nontrivial terms. As noted in \cite{Damanik:Sims:Stolz:1} the set of zeroes of the
Lyapunov exponent is contained in the union
\begin{equation*}
\bigcup_{\substack{ s\neq s' \\ s,s'\in\supp\varkappa}} S(s,s')
\end{equation*}
of the sets \eqref{das:S}. The Lyapunov exponent is strictly positive for all $E$ away from this
discrete set.

\begin{figure}[htb]
\ifpdf \centerline{\includegraphics[width=120mm]{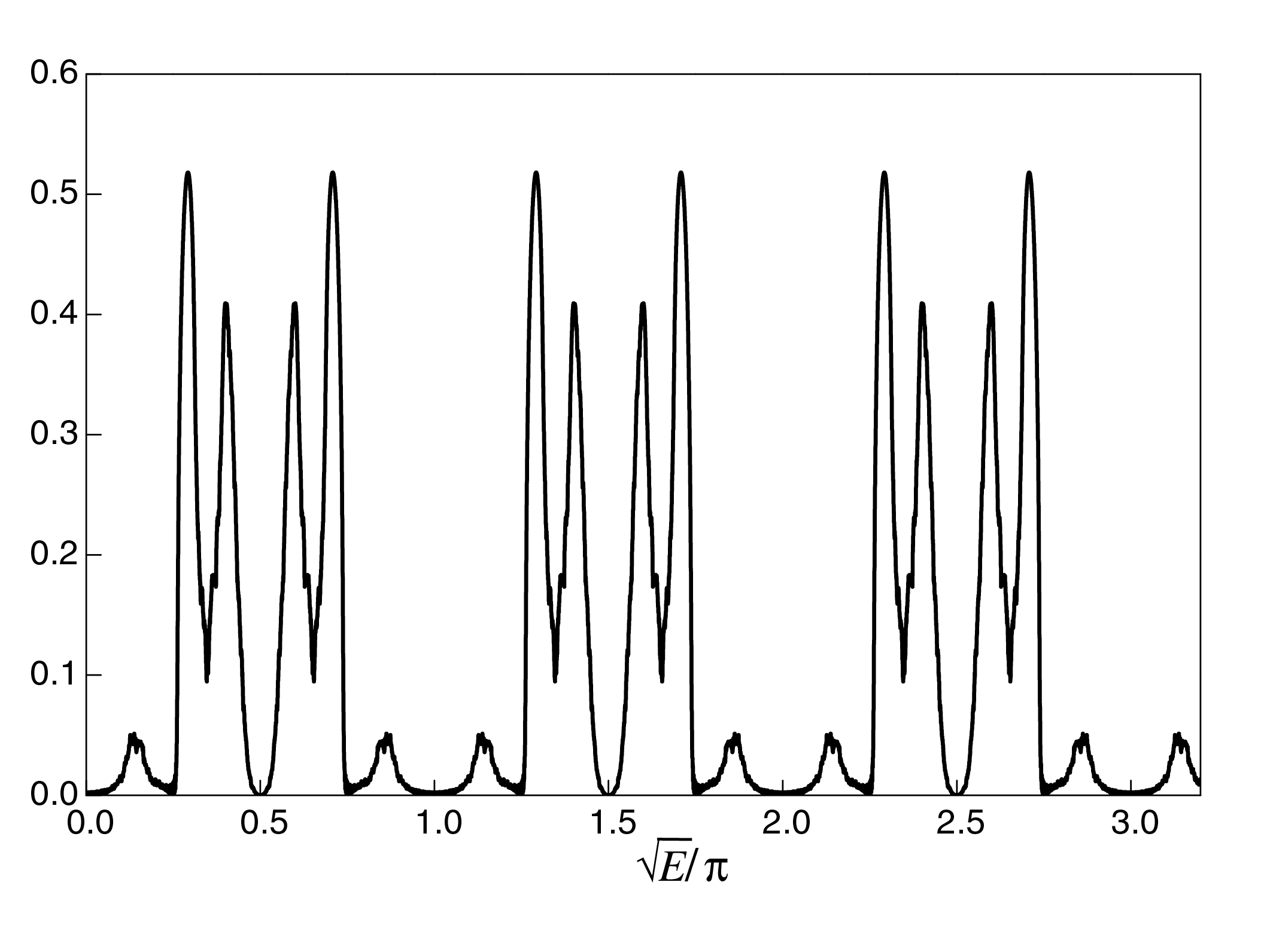}} \else
\vspace{47mm}\centerline{Bernoulli\_gam.png} \vspace{47mm}\fi
\caption{\label{fig:ring:bin} The Lyapunov exponent for the Bernoulli
distribution $\varkappa= \frac{1}{2}\delta_2 + \frac{1}{2}\delta_6$, i.e.,
$\omega_j\in\{2,6\}$ with equal probability.}
\end{figure}

Results of numerical computations of the Lyapunov exponent using \eqref{multip:gam} for the
Bernoulli distribution with $\varkappa= \frac{1}{2}\delta_2 + \frac{1}{2}\delta_6$, i.e.,
$\omega_j\in\{2,6\}$ with equal probability, are presented in Figure \ref{fig:ring:bin}. We
mention two properties of $\gamma(E)$:

1. The Lyapunov exponent is periodic in $\sqrt{E}$ with period $\pi$, i.e.\
\begin{equation*}
\gamma((\sqrt{E}+\pi k)^2)=\gamma(E),\quad E>0,\quad k\in\N.
\end{equation*}
This follows immediately from the fact that for even $s$
\begin{equation*}
{\Lambda}_s((\sqrt{E+\pi k})^2)={\Lambda}_s(E).
\end{equation*}

2. The Lyapunov exponent is reflection symmetric,
\begin{equation}\label{Lyapunov.reflect}
\gamma((k\pi-\sqrt{E})^2)=\gamma(E),\quad E>0,\quad k\in\N,\quad
k>\sqrt{E}/\pi,
\end{equation}
i.e., the points $k/2$, $k\in\N$ are the axes of reflection symmetry (on the
scale of $\sqrt{E}/\pi$). To prove this we note that
\begin{equation*}
\begin{split}
{\Lambda}_s((k\pi-\sqrt{E})^2)=& \begin{pmatrix}
-\frac{e^{-2is\sqrt{E}}-9}{8e^{-is\sqrt{E}}}e^{i\sqrt{E}}e^{-ik\pi} &
\frac{3i}{4}\sin(s\sqrt{E})\\
-\frac{3i}{4}\sin(s\sqrt{E}) &
-\frac{e^{2is\sqrt{E}}-9}{8e^{is\sqrt{E}}}e^{-i\sqrt{E}}e^{ik\pi}
\end{pmatrix}\\
 =& \begin{pmatrix}
     0 & e^{ik\pi}\\
     1 & 0 \end{pmatrix} {\Lambda}_s(E)
     \begin{pmatrix}
     0 & e^{ik\pi}\\
     1 & 0 \end{pmatrix}.
\end{split}
\end{equation*}
Since
\begin{equation*}
\begin{pmatrix}
     0 & e^{ik\pi}\\
     1 & 0 \end{pmatrix}^2=e^{ik\pi} \begin{pmatrix}
     1 & 0\\
     0 & 1 \end{pmatrix},
\end{equation*}
up to a sign the product $\prod_{j=-n}^m
{\Lambda}_{\omega_j}((k\pi-\sqrt{E})^2)$  equals
\begin{equation*}
\begin{pmatrix}
     0 & e^{ik\pi}\\
     1 & 0 \end{pmatrix} \prod_{j=-n}^m{\Lambda}_{\omega_j}(E)
\begin{pmatrix}
     0 & e^{ik\pi}\\
     1 & 0 \end{pmatrix}.
\end{equation*}
Thus, equality \eqref{Lyapunov.reflect} follows from \eqref{multip:gam}.

Using equation \eqref{multip:gam} one can also analyze the periodic case
$\omega_j=\omega_0$ for all $j\in\Z$. In this case the spectrum of
$-\Delta(\omega)$ consists of the absolutely continuous part and the
eigenvalues $E_k=\pi^2 k^2/\omega_0^2$, $k\in\N$ of infinite multiplicity.
The absolutely continuous spectrum has a band structure such that
$E\in\specac(-\Delta(\omega))$ if and only if Hill's discriminant (cf.\
equation (8) in \cite{Avron:Exner:Last})
\begin{equation*}
H(E)=\frac{2\cos(\sqrt{E}+\phi_{0,0}(E;\omega))}{|T_{0,0}(E;\omega)|}=
\frac{9}{4}\cos(\sqrt{E}(\omega_0+1))-\frac{1}{4}\cos(\sqrt{E}(\omega_0-1))
\end{equation*}
satisfies the inequality $|H(E)|\leq 2$. {}From this and the fact that
$|T_{0,0}(\pi^2n^2/a^2;\omega)|=1$ it follows immediately that all
eigenvalues are imbedded in the absolutely continuous spectrum or lie at the
edges of the spectral bands.

\section{Continuity of $\widetilde{N}(E)$}\label{cont:IDS}

For almost all $E>0$ the Lyapunov exponent satisfies the Thouless formula
\begin{equation}\label{Thouless}
\gamma(E) = \alpha + \int_\R \log\left|\frac{\lambda-E}{\lambda-i}\right|
d\widetilde{N}(\lambda),
\end{equation}
where $\alpha$ is some positive number. The existence of the integral on the
r.h.s.\ is guaranteed by the estimate \eqref{estimate} and Lemma 11.7 in
\cite{Pastur:Figotin}. We emphasize that $N^{\mathrm{loop}}(E)$ does not
enter this formula.

Before we proceed, we discuss the implications of \eqref{Thouless} to the
continuity properties of $\widetilde{N}(E)$. By a modification of an
argument due to Craig and Simon \cite{Craig:Simon} (see Theorem 11.9 in
\cite{Pastur:Figotin}) the positivity of the Lyapunov exponent implies the
log-H\"{o}lder continuity of $\widetilde{N}(E)$, that is, the inequality
\begin{equation*}
\widetilde{N}(E_2) - \widetilde{N}(E_1) \leq C
\left|\log|E_2-E_1|\right|^{-1}
\end{equation*}
for arbitrary sufficiently small intervals $[E_1, E_2]$. Moreover, using the
arguments (in a slightly modified form) of Damanik, Sims, and Stolz from
\cite{Damanik:Sims:Stolz} one proves that $\widetilde{N}(E)$ is actually
H\"{o}lder continuous, i.e., there is a number $0<\mu<1$ such that
\begin{equation*}
\widetilde{N}(E_2) - \widetilde{N}(E_1) \leq C |E_2-E_1|^\mu
\end{equation*}
for arbitrary sufficiently small intervals $[E_1, E_2]$ not containing the points where the
Lyapunov exponent vanishes. Both these properties hold for general distribution measures
$\varkappa$.

We only sketch the proof of \eqref{Thouless}, but the interested reader can easily fill in the
details. We closely follow the line of arguments given in \cite{Kostrykin:Schrader:1999a}.

Let $f(z)$ be an analytic function in open cut plane $\C_0=\C\setminus
[0,\infty)$, continuous in the closure of $\C_0$ and satisfying $|f(z)|\leq
c \sqrt{|z|}$ for all $z\in\C_0$. Moreover, we assume that
$f(E+i0)=\overline{f(E-i0)}$ is continuously differentiable for all $E>0$.
Using the Cauchy integral formula it is easy to show that
\begin{equation*}
\Re f(z) = \Re f(i) +\frac{1}{\pi} \int_\R
\log\left|\frac{\lambda-z}{\lambda-i}\right| d\ \Im f(\lambda+i0).
\end{equation*}
{}From equations \eqref{T.R.L.} and by the factorization rule for the
matrices \eqref{Lambda:def} (see \cite{Kostrykin:Schrader:2001b}) it
follows that one can take $f(z) = \log T_{m,n}(z;\omega)$ for arbitrary
$m$, $n$, and $\omega\in\Omega$, thus, obtaining
\begin{equation*}
\log|T_{m,n}(z;\omega)| = \log|T_{m,n}(i;\omega)| +\frac{1}{\pi} \int_\R
\log\left|\frac{\lambda-z}{\lambda-i}\right| d\ \phi_{m,n}(\lambda;\omega).
\end{equation*}
Using Lemma 11.7 in \cite{Pastur:Figotin} we conclude that for almost all $E>0$, all
$\omega\in\Omega$, and arbitrary integers $m$, $n$
\begin{equation*}
\log|T_{m,n}(E;\omega)| = \log|T_{m,n}(i;\omega)| +\frac{1}{\pi} \int_\R
\log\left|\frac{\lambda-E}{\lambda-i}\right| d\ \phi_{m,n}(\lambda;\omega).
\end{equation*}
Now divide both sides of this equation by $n+m+1$ and consider the limit
$m,n\rightarrow\infty$. By \eqref{Lyapunov:2} the l.h.s.\ converges almost
surely to $-\gamma(E)$. The signed measures $(n+m+1)^{-1}d\
\phi_{m,n}(E;\omega)$ converge vaguely to $\pi
(dN_0(E)-d\widetilde{N}(E))$. Thus, again by Theorem 11.7 in
\cite{Pastur:Figotin} there are subsequences $m_k,n_k$ such that
\begin{equation*}
\lim_{k\rightarrow\infty}\frac{1}{\pi}\int_\R
\log\left|\frac{\lambda-E}{\lambda-i}\right| \frac{d\
\phi_{m_k,n_k}(\lambda;\omega)}{m_k+n_k+1} = \int_\R
\log\left|\frac{\lambda-E}{\lambda-i}\right| (d N_0(\lambda)-dN(\lambda))
\end{equation*}
for almost all $E$. Noting that
\begin{equation*}
\int_\R \log\left|\frac{\lambda-E}{\lambda-i}\right| d N_0(\lambda) =
\gamma_0(E)-\gamma_0(i)= -\frac{\sqrt{2}}{2},
\end{equation*}
where $\gamma_0(z)=|\Re\sqrt{-z}|$ is the Lyapunov exponent of the
Laplacian $-\Delta$ on $L^2(\R)$, we obtain \eqref{Thouless}.

\appendix
\section{The Spectral Shift Function}\label{Appendix:A}

Here we briefly collect some facts from the theory of the spectral shift function in the case
where the operators involved act in different Hilbert spaces. For a comprehensive presentation we
refer the reader to the book \cite{Yafaev}. Consider two (possibly unbounded) self-adjoint
operators $T_0\geq I$ and $T\geq I$ acting in Hilbert spaces $\cH_0$ and $\cH$, respectively, and
a bounded operator $\cJ:\ \cH_0\rightarrow\cH$. Suppose that the operators
\begin{equation}\label{Koplienko}
\left.\begin{array}{l} T^{-1}\cJ-\cJ T_0^{-1}\\ (\cJ^\ast \cJ-I)T_0^{-1}\\
T^{-1}(\cJ\cJ^\ast-I)\end{array}\right\}\qquad \textrm{are trace class}.
\end{equation}
Under these conditions there exists a spectral shift function
$\xi(E;T,T_0;\cJ)$ for which the trace formula
\begin{equation}\label{Koplienko.trace}
\tr\left[\phi(T)-\cJ\phi(T_0)\cJ^\ast
\right]+\tr\left[(\cJ^\ast\cJ-I)\phi(T_0)\right]= \int_\R \xi(E;T,
T_0;\cJ)\phi^\prime(E)dE
\end{equation}
holds, where $\phi$ is an arbitrary bounded continuously differentiable complex-valued function.
The relation to the scattering matrix is given by the Birman-Krein theorem
\begin{equation}\label{Appendix:Birman:Krein}
\det S(E;T,T_0;\cJ)=\exp\{-2\pi i\xi(E;T,T_0;\cJ)\},
\end{equation}
where $S(E;T,T_0;\cJ)$ is the scattering matrix for the operators $(T,T_0)$.


\begin{thebibliography}{50}

\bibitem{Arai:Tokihiro:Fujiwara:Kohmoto} M.~Arai, T.~Tokihiro, T.~Fujiwara, and M.~Kohmoto,
\textit{Strictly localized states on a two-dimensional Penrose lattice},
Phys. Rev. B \textbf{38}, 1621 -- 1626 (1988).

\bibitem{Avron:Exner:Last} J.~Avron, P.~Exner, and Y.~Last, \textit{Periodic Schr\"{o}dinger
operators with large gaps and Wannier-Stark ladders}, Phys. Rev. Lett. \textbf{72}, 896 -- 899
(1994).

\bibitem{Carmona:Lacroix} R.~Carmona and J.~Lacroix, \textit{Spectral Theory of Random
Schr\"{o}dinger Operators}, Boston, Birkh\"{a}user, 1990.

\bibitem{Craig:Simon} W.~Craig and B.~Simon, \textit{Log H\"{o}lder continuity of the integrated
density of states for stochastic Jacobi matrices}, Comm. Math. Phys.
\textbf{90}, 207 -- 218 (1983).

\bibitem{Damanik:Sims:Stolz:1} D.~Damanik, R.~Sims, and G.~Stolz,
\textit{Lyapunov exponents in continuum Bernoulli-Anderson models}, in S.~Albeverio, N.~Elander,
W.~N.~Everitt, and P.~Kurasov (eds.), \textit{Operator Methods in Ordinary and Partial
Differential Equations} (Stockholm, 2000), p. 121 -- 130, Oper. Theory Adv. Appl., Vol. 132,
Birkh\"{a}user, Basel, 2002.

\bibitem{Damanik:Sims:Stolz} D.~Damanik, R.~Sims, and G.~Stolz, \textit{Localization for
one dimensional, continuum, Bernoulli-Anderson models}, Duke Math. J.
\textbf{114}, 59 -- 100 (2002).

\bibitem{Fujiwara:Arai:Tokihiro:Kohmoto} T.~Fujiwara, M.~Arai, T.~Tokihiro, and M.~Kohmoto,
\textit{Localized states and self-similar states of electrons on a two-dimensional Penrose
lattice}, Phys. Rev. B \textbf{37}, 2797 -- 2804 (1988).

\bibitem{Kirsch:Martinelli} W.~Kirsch and F.~Martinelli, \textit{On the density of
states of Schr\"{o}dinger operators with a random potential}, J. Phys. A: Math. Gen. \textbf{15}, 2139
-- 2156 (1982).

\bibitem{Klassert:Lenz:Stollmann} S.~Klassert, D.~Lenz, and P.~Stollmann, \textit{Discontinuities of the
integrated density of states for random operators on Delone sets}, Comm.
Math. Phys. (to appear); DOI:\ 10.1007/s00220-003-0920-7;
\texttt{arXiv:math-ph/0208027}.

\bibitem{Kohmoto:Sutherland} M.~Kohmoto and B.~Sutherland, \textit{Electronic states on a Penrose
lattice}, Phys. Rev. Lett. \textbf{56}, 2740 -- 2743 (1986).

\bibitem{habil} V.~Kostrykin, \textit{The spectral shift function and its applications to
random Schr\"{o}dinger operators}, Habilitationsschrift, RWTH Aachen, 1999.

\bibitem{Kostrykin:Schrader:1998} V.~Kostrykin and R.~Schrader, \textit{One-dimensional disordered systems
and scattering theory}, Sfb 288 Preprint 337, Berlin, 1998. Available
from\newline \texttt{http://www-sfb288.math.tu-berlin.de/abstractNew/337}.

\bibitem{Kostrykin:Schrader:1999a} V.~Kostrykin and R.~Schrader, \textit{Scattering theory
approach to random Schr\"{o}dinger operators in one dimension}, Rev. Math. Phys.
\textbf{11}, 187 -- 242 (1999).

\bibitem{Kostrykin:Schrader:1999b} V.~Kostrykin and R.~Schrader,
\textit{Kirchhoff's rule for quantum wires}, J. Phys. A: Math. Gen. \textbf{32}, 595 -- 630
(1999).

\bibitem{Kostrykin:Schrader:2000b} V.~Kostrykin and R.~Schrader,
\textit{Global bounds for the Lyapunov exponent and the integrated density of states of random
Schr\"{o}dinger operators in one dimension}, J. Phys. A: Math. Gen. \textbf{33}, 8231 -- 8240 (2000).

\bibitem{Kostrykin:Schrader:2001b} V.~Kostrykin and R.~Schrader,
\textit{The generalized star product and the factorization of scattering
matrices on graphs}, J. Math. Phys. \textbf{42}, 1563 -- 1598 (2001).

\bibitem{Kostrykin:Schrader:2003} V.~Kostrykin and R.~Schrader,
\textit{Quantum wires with magnetic fluxes}, Comm. Math. Phys. \textbf{237}, 161 -- 179 (2003).

\bibitem{Krajci:Fujiwara} M.~Kraj{\fontencoding{T1}\selectfont\symbol{'243}\symbol{'355}}
and T.~Fujiwara, \textit{Strictly localized eigenstates on a
three-dimensional Penrose lattice}, Phys. Rev. B \textbf{38}, 12903 --
12907 (1988).

\bibitem{Lifshitz:Gredeskul:Pastur:82} I.~M.~Lifshitz, S.~A.~Gredeskul, and
L.~A.~Pastur, \textit{Theoty of the passage of particles and waves through
randomly inhomogeneous media}, Sov. Phys. JETP \textbf{56}, 1370 -- 1378
(1982).

\bibitem{Lifshitz:Gredeskul:Pastur:88} I.~M.~Lifshitz, S.~A.~Gredeskul, and
L.~A.~Pastur, \textit{Introduction to the Theory of Disordered Systems},
Wiley, New York, 1988.

\bibitem{Marchenko:Pastur} A.~V.~Marchenko and L.~A.~Pastur, \textit{Transmission of waves and
particles through long random barriers}, Theor. Math. Phys. \textbf{68}, 929
-- 940 (1986).

\bibitem{Pastur:Figotin} L.~Pastur and A.~Figotin, \textit{Spectra of Random and Almost-Periodic
Operators}, Springer, Berlin, 1992.

\bibitem{Yafaev} D.~R.~Yafaev, \textit{Mathematical Scattering Theory. General Theory},
Amer. Math. Soc. Transl. of Math. Monographs Vol. 105, Providence, RI, 1992.

\end{thebibliography}
\end{document}